\documentclass{pap}

\usepackage{psfig,epsfig}
\usepackage{graphicx,latexsym}
\usepackage{psfrag,amsmath}
\usepackage[english]{babel}

\newcommand{\tB}{{\text{B}}}

\newcommand{\gp}{\sigma_{pr}}
\newcommand{\de}{\text{d}}
\newcommand{\p}[1]{\left({#1}\right)}
\newcommand{\pq}[1]{\left[{#1}\right]}

\title{Shape fluctuations and elastic properties of two-component bilayer membranes}
\author{Alberto Imparato\inst{1} \thanks{Present address:
Dipartimento di Scienze Fisiche and Unit\`a INFM,
Universit\`a ``Federico II'', Complesso Universitario di Monte S. Angelo,
I-80126 Napoli (Italy).}
 \and Julian C. Shillcock \inst{1} \and Reinhard  Lipowsky\inst{1}}
\institute{
\inst{1}{Max-Planck-Institut f\"ur Kolloid- und
Grenzfl\"achenforschung,\\ 14424 Potsdam, Germany.\\
}
}

\begin{document}

\maketitle

\begin{abstract}
The elastic properties of  two-component bilayer membranes
are studied using a coarse grain model for amphiphilic
molecules. The two species of amphiphiles considered here differ only in
their length.
Molecular Dynamics simulations are performed in order to
analyze the shape fluctuations of the two-component bilayer membranes
 and to determine their bending rigidity.
Both the bending rigidity and its inverse are found to be nonmonotonic 
functions  of the mole fraction  $x_{\rm B}$  of the shorter
B-amphiphiles and, thus, do not satisfy a simple lever rule.
The intrinsic area of the bilayer also exhibits
a nonmonotonic dependence on $x_{\rm B}$ and
a maximum close to $x_{\rm B} \simeq 1/2$.

\end{abstract}

Biological membranes are multicomponent systems consisting of
mixtures of many different lipids and proteins. Because the
composition of lipid bilayers affects their
physical properties and biological functions, this
composition varies from one organism to the other and between organelles
in the same cell \cite{LS}.
Biological membranes contain a fluid bilayer which is highly flexible
and, thus, can easily change its shape. Typical examples are
provided by the plasma membranes of red and white
blood cells which are so flexible that they can move
through rather small capillaries.
This flexibility is also responsible for the  thermally excited
shape fluctuations of  biomembranes in physiological
conditions, the amplitude of which depends on temperature,  membrane
composition
and  mechanical constraints.
One important example is provided by mixed membranes containing phospholipids
and cholesterol which exhibit a strong increase in rigidity with increasing
amount of cholesterol  \cite{DKS,ER,MG}.
In erythrocytes, on the other hand,
the amplitude of the shape fluctuations is influenced by the spectrin-ankyrin
network which is coupled to the interior of the plasma membrane \cite{WE,SPS}.
The study of these  fluctuations can thus  provide
 information on the elastic properties of the bilayer membrane and
on how these properties depend on  membrane composition.
One important elastic parameter is the  {\it bending rigidity}
\cite{canh70}, which  describes
the resistance of the membrane to bending,  and which can
be obtained from the spectrum of the shape fluctuations \cite{LB}.

In spite of their complex chemical composition, all biomembranes have the same
basic structure: a bilayer of lipid and protein molecules. Therefore the
simplest
 model for a biological membrane is a bilayer
of amphiphilic molecules as studied  in the present work.
The bending rigidity of  such bilayers has already been determined
via computer simulations both for coarse grain models of amphiphile-water
systems \cite{GL2,MVM} and for models with atomic resolution \cite{LE1}.
In all of these previous studies, the bilayers contained only
a single component in contrast to
natural membranes which  contain many different
types of lipids and proteins.

In the present article, we study bilayer membranes with two components
and determine the dependence of their elastic properties on membrane
composition. We
use a  coarse grain model for the amphiphiles and for the water particles
which we investigate by Molecular Dynamics simulations.
The two species of amphiphiles considered here differ only in  their length.
This choice allows us to focus  on the contribution from the mismatch of
the amphiphile tails.

Two types of theories have previously addressed the 
bending elasticity of two--component bilayers. First, a 
self--consistent (or molecular field) theory was used
to estimate the  contribution to the bending rigidity
arising from the conformations of the amphiphilic 
tails \cite{szkr}.  Secondly, the
so--called hat model for local curvature fluctuations 
was used in order to derive an expression for the 
bending rigidity of two--component bilayers \cite{Marki}. 
The predictions of these two theories are 
different as far as the functional dependence of
 bending rigidity on composition is concerned. 
On the one hand, the results of the self--consistent
theory for the chain conformations  depend  on the
 types of packing constraint  acting on the 
conformations of a single tail.  The hat model, on
the other hand, predicts that the inverse bending 
rigidity (or flexibility) satisfies a simple `lever 
rule' corresponding to a linear interpolation between 
the two limiting values for the one--component bilayers. 

In contrast to the analytical but approximate theories
in \cite{szkr} and \cite{Marki}, our simulations take  all types of fluctuations
into account: different
 tail conformations; curvature (or bending)
fluctuations; molecular protrusions; and composition fluctuations
arising from lateral diffusion. We find that both the bending
rigidity and its inverse exhibit a nonmonotonous behavior
as a function of the mole fraction $x_{\rm B}$. Our results
are qualitatively similar to the self--consistent theory, 
{\em provided} one uses the packing constraint of `constant area' 
per molecule, but show that a simple lever rule for the inverse
bending rigidity does not apply in general. 

\begin{figure}[h]
\center
\psfrag{A}[b][Bl]{A}
\psfrag{B}[b][Bl]{B}
\includegraphics[height=2.4cm,width=7.cm]{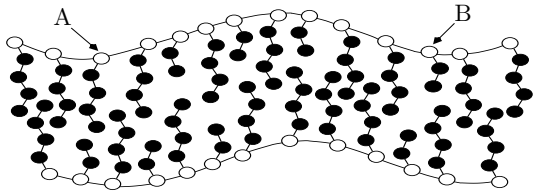}
\caption{Cartoon of a two-component bilayer.}
\label{byl}
\end{figure}

The coarse grain model used here consists of several types of
particles: water particles, which correspond to several water molecules,
as well as head group and tail particles, which represent the head groups
and the hydrocarbon tails of the amphiphilic molecules, respectively.
The two species of amphiphiles will be denoted by A and B.
Both types of amphiphiles are composed of one head group
particle, which is hydrophilic, and of several tail particles, which
are hydrophobic. The different length of the two molecular species
arises from the different number of tail particles which is four and two
for A and B, respectively, see figure \ref{byl}.
The particles experience  attractive and repulsive pair-interactions
which are modelled
by Lennard-Jones (LJ) and  Soft-Core potential (SC), respectively.
The parametrization of these potentials has been previously described
in \cite{alb1} where the same  model was used in order
to study the lateral and transverse diffusion within the two-component
bilayer membranes. Within this model, all
quantities can be expressed in terms of three basic scales:
the particles mass  $m$, the LJ interaction energy $\epsilon$
and the LJ radius $\rho$. As in \cite{alb1} and \cite{GL1},
we chose the numerical values
$m N_{Av}=3.6\cdot 10^{-2}$kg, $\epsilon N_{Av}=2$kJ, and $\rho=1/3$ nm,
where $N_{Av}$ is the Avogadro number.
Our MD simulations are carried out for a cuboidal box of constant volume
and periodic boundary conditions
using a constant temperature algorithm with $k_B T=1.35 \epsilon$.
The starting configurations for the two-component bilayer simulations
are such that each monolayer contains the same number of A and B molecules.
We checked that, in each monolayer, the average number
of B molecules did not change during the simulations.
While multicomponent bilayer membranes may exhibit deviations from 
ideal mixing behavior \cite{nonid} and may, in general, undergo phase separation, the 
two--component bilayers studied here stayed in the one--phase region 
over the whole range
$0 \le  x_{\rm B}  \le 1$ of mole fractions for the B molecules. Thus, our 
bilayers correspond to a lipid mixture above the liquidus--solidus 
line.

The elastic properties of a fluid membrane which has vanishing
spontaneous curvature are governed
by its bending rigidity $\kappa$ and by its surface tension $\sigma$.
These quantities can be determined from the spectrum of the shape
fluctuations. In order to do so, we choose the $(x,y)$-plane to
be parallel to the bilayer membrane.  The shape of the membrane is
then described by the height function $h(x,y)$  which measures the
distance of its midsurface from the $(x,y)$-plane. We  define
the Fourier coefficients
$\tilde h_{\mathbf{q}} \equiv  A_p^{-1}
\int_{A_p} dx dy \, \exp\pq{-i(x q_x +y q_y)} h(x,y)$,
with $\mathbf{q} \equiv (q_x,q_y)$, where $A_p$ is the
projected area of the membrane. Its fluctuation spectrum,
$ S(q)\equiv \langle \, |\tilde h_{\mathbf{q}}|^2 \rangle$,
depends only on $q = |{\mathbf{q}}|$ and
exhibits the functional form
\begin{equation}
S(q)  = k_B T/\pq{A_p\p{\sigma q^2+ \kappa q^4}} 
\label{sq1}
\end{equation} 
for long-wavelength bending modes  \cite{LB},
and the somewhat different form
\begin{equation}
S(q)  = k_B T/\p{A_p\sigma_{pr} q^2}  
\label{sq2}
\end{equation} 
for short-wavelength molecular protrusions \cite{LG}.
Apart from the temperature $T$ and the projected area $A_p$,
the two spectra as given by (\ref{sq1}) and (\ref{sq2})
contain three parameters: the  surface tension $\sigma$,
the bending rigidity $\kappa$, and the protrusion tension
$\sigma_{pr}$.
As discussed in \cite{GL1}, the lateral size of the simulation box can be
adjusted in order to obtain a bilayer with  vanishing thermodynamic tension. 
This latter procedure consists in determining, for fixed particle 
number, fixed volume, and fixed temperature, the stress (or pressure)
tensor which has a tangential component, $\Sigma_T$, and a
normal component, $\Sigma_N$ \cite{GL1}.
Both components depend on the coordinate $z$
perpendicular to the membrane. The thermodynamic tension $\Sigma$
is then obtained from $\Sigma = \int  dz [\Sigma_T(z) - \Sigma_N (z)]$.
It is intuitively appealing to identify the thermodynamic tension
$\Sigma$ with the surface tension $\sigma$ which governs the
fluctuation spectrum. Indeed, this assumption was implicit in our
previous work in \cite{GL2} and was found to be satisfied
within the accuracy of the simulation data obtained there.
However, more extensive simulations have shown
that this identity does not hold in general \cite{alb2}.
This difference is related to the fact that $\Sigma$ is
conjugate to the projected area, whereas $\sigma$ is
conjugate to the intrinsic area, a distinction which has
been previously discussed, e.g., in \cite{DL}.
Therefore, in the present work, we used an improved procedure
in order to determine the state with $\sigma = 0$: for fixed particle
number and fixed temperature, we
 adjusted the lateral box size in
order to obtain a fluctuation spectrum $S(q)$ which can
be well fitted, for small $q$,  by the form
 (\ref{sq1}) with $\sigma = 0$ \cite{alb2}.

\begin{figure}[h]
\center
\psfrag{S}[cl][cl][1.]{$S\, \pq{\rho^2}$}
\psfrag{q}[cl][cl][1.]{$q\, \pq{\rho^{-1}}$}
\psfrag{x0}[rc][rc][.8]{$x_{\rm B}=0$}
\psfrag{x1}[rc][rc][.8]{$x_{\rm B}=0.4$}
\psfrag{l1}[rc][rc][.8]{$\sim 1/(\kappa q^4)$}
\psfrag{l2}[rc][rc][.8]{$\sim 1/(\gp q^2)$}
\includegraphics[height=5.5cm,width=8cm]{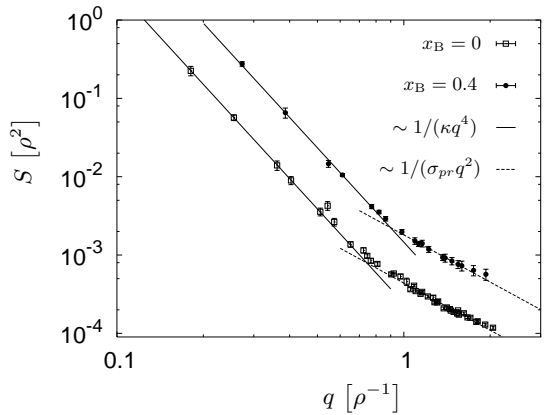}
\caption{Fluctuation spectrum $S$ for   one and two-component bilayers with 
$N_{\rm A}=1152$ ($x_{\rm B}=0$) and $N=512$ ($x_{\rm B}=0.4$), respectively,
in  tension-free state. The lines are obtained by fitting
the fluctuation spectra of the tensionless systems, using eq. (\ref{sq1}) (full lines) and  eq. (\ref{sq2}) (dashed lines) at small  and large $q$, respectively. The units are given in brackets.}
\label{sq_gamma_0}
\end{figure}

First, let us discuss this fitting procedure for
one-component bilayers  corresponding to vanishing mole fraction
$x_{\rm B} = 0$. An example of the measured fluctuation spectrum $S(q)$
for $N_{\rm A}=1152$.
is shown in figure \ref{sq_gamma_0}.
The lines, which represent  the functional forms as given
by (\ref{sq1}) with $\sigma = 0$ and  by (\ref{sq2}), respectively, are 
 plotted in the same figure.
From these fits, we obtain the values $\kappa = 3.0 \pm 0.2\,  k_B T$ and
$\sigma_{pr} =2.50 \pm 0.03\,  \epsilon/\rho^2$. Essentially the same values are found for
$N_{\rm A} = 512$: $\kappa = 3.20 \pm 0.06\,  k_B T$ and
$\sigma_{pr} = 2.65 \pm 0.05\,  \epsilon/\rho^2$ \cite{alb2}.
For surface tension $\sigma = 0$, the intrinsic area per molecule, $a_{in}$, 
is expected to attain a certain  value which is determined by the optimal
packing of the molecules within the bilayer and, thus, should not depend
on the overall size of the membrane. Indeed, if we adjust the
box size to obtain vanishing surface tension $\sigma \simeq 0$,
our MD simulations lead to $a_{in} = 2.26\pm 0.01\, \rho^2$
for $N_{\rm A} = 512$ and to  $a_{in} =2.277\pm 0.005\, \rho^2$ for
$N_{\rm A} = 1152$ which are identical within the numerical accuracy.

Next, we consider two-component bilayers with mole fraction $0 < x_{\rm B} <1$.
To save computation time, we  varied the mole fraction $x_{\rm B}$ for
fixed total number 
$N=N_{\rm A}+N_{\rm B}$ of the amphiphiles with $N=512$.
 In order to attain tensionless states of the  two-component bilayers,
the projected area $A_p$ has to be gradually decreased as  $x_{\rm B}$ 
is increased,
since the  projected area per molecule, $a_p\equiv 2 A_p/N$, is smaller
 for the  B-amphiphile than for the A--amphiphile.
The measured fluctuation spectrum $S(q)$ for $x_{\rm B}=0.4$ is also shown
in figure \ref{sq_gamma_0}.

\begin{figure}[h]
\center
\psfrag{kappa}[br][br][1.]{$\kappa\,  \pq{k_BT}$}
\psfrag{xb}[ct][ct][1.]{$x_\tB$}
\psfrag{fa}[bl][bl][1.]{$(a)$}
\psfrag{1/k}[tr][tr][1.]{$1/\kappa\,  \pq{(k_BT)^{-1}}$}
\psfrag{fb}[bl][bl][1.]{$(b)$}
\includegraphics[width=7.cm]{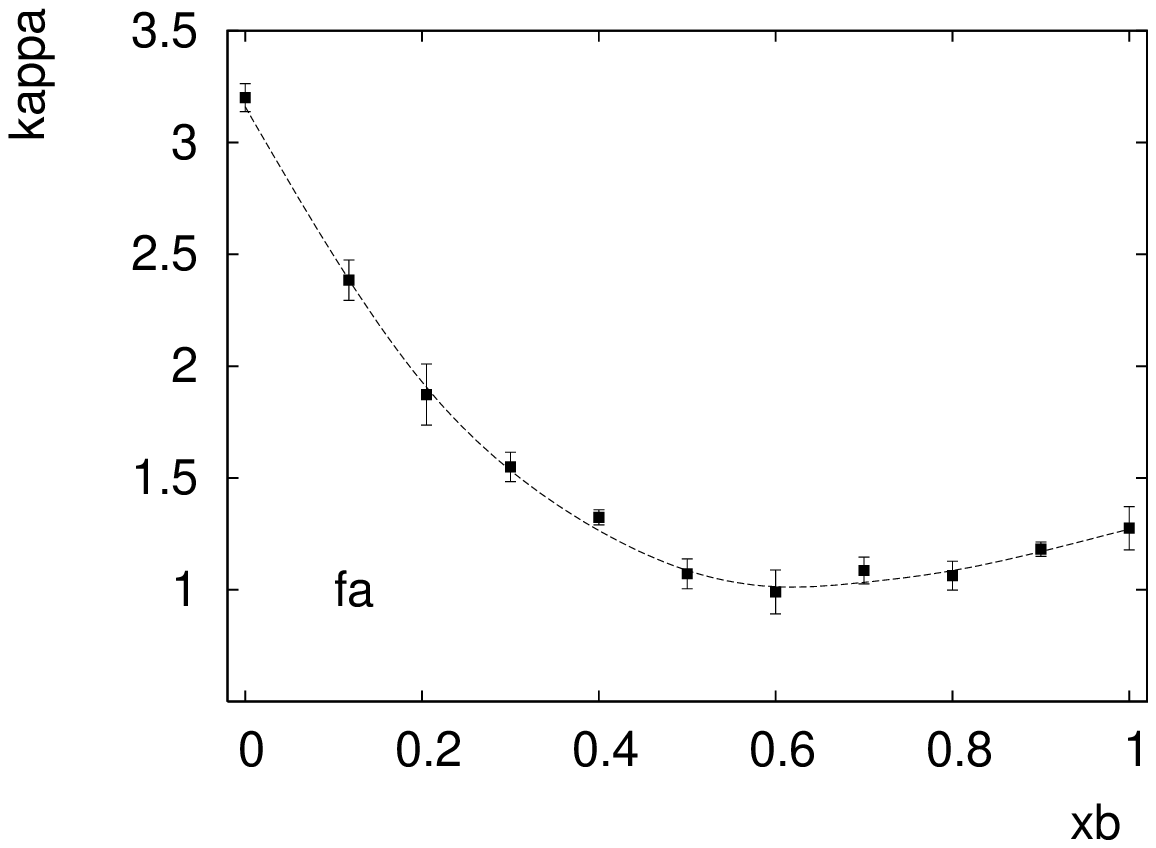}
\includegraphics[width=7.cm]{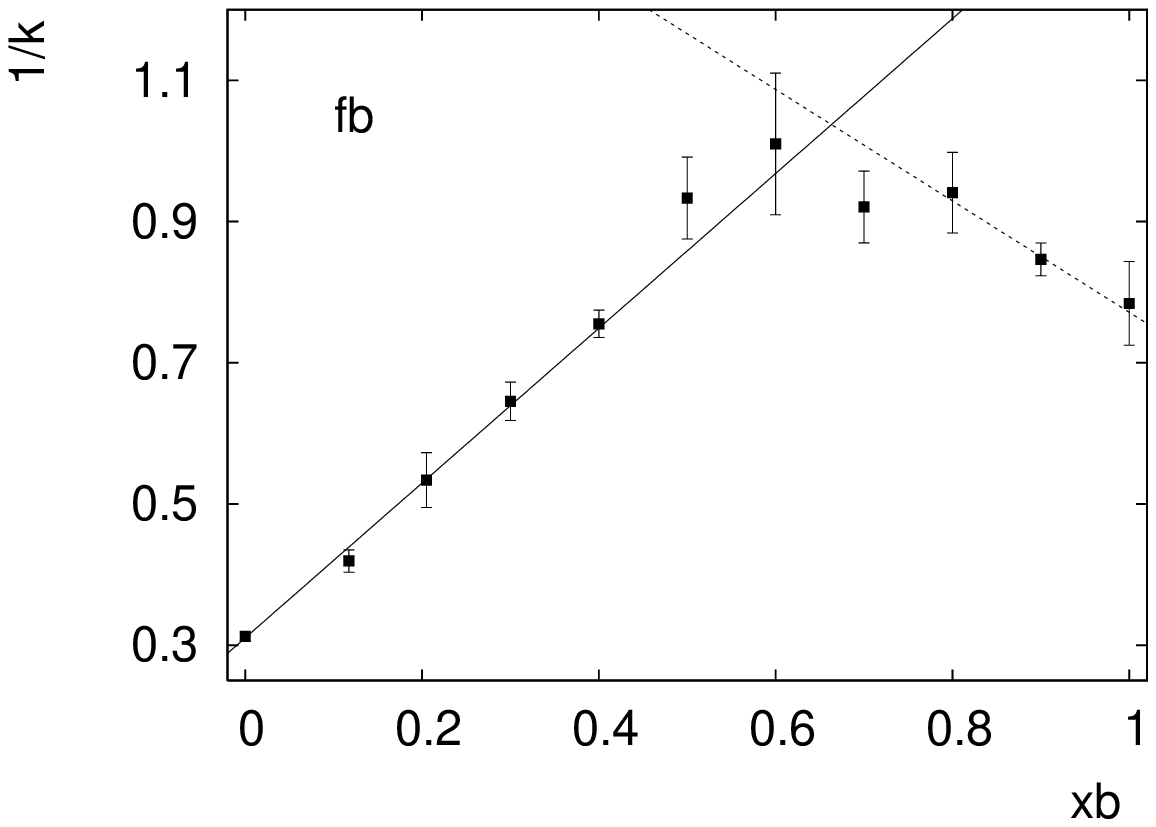}
\caption{(a) Bending rigidity $\kappa$ and (b) inverse bending rigidity of  two-component A- B-bilayers vs. the shorter amphiphile mole concentration $x_\tB$. The error bars represent statistical
errors obtained by the fit of the fluctuation spectrum data  to eq. (\ref{sq1}).
In (a)  the line
is  obtained by fitting the data with a smoothing spline. In (b) the full and dotted line correspond to linear fit
to the data  for $x_\tB\leq 0.4$ and $x_\tB\geq 0.8$, respectively. The units are given in brackets.}
\label{usk}
\end{figure}

In figure
\ref{usk}(a), the bending rigidity $\kappa$ is plotted  as a function
of  $x_{\rm B}$.
For the pure A--bilayer and the pure B-bilayer, we find
$\kappa_{\rm A} = 3.20  \pm 0.06\, k_B T$ and $\kappa_{\rm B} = 1.3 \pm 0.1\, k_B T$,
respectively. As shown in figure \ref{usk}(a), the bending rigidity is
nonmonotonic for intermediate values of  $x_{\rm B}$
and exhibits a  minimum for $x_{\rm B}\simeq 0.6$  with
$\kappa = 1.0 \pm 0.1\, k_B T$.
Figure \ref{usk}(b) displays the inverse bending rigidity $1/\kappa$
as a function of the mole fraction $x_{\rm B}$.
For this latter quantity,  a simple lever rule as given by $1/\kappa =
x_{\rm B}/\kappa_{\rm B} + (1-x_{\rm B})/\kappa_{\rm A}$ has been proposed in
\cite{Marki}.
Inspection of figure \ref{usk}(b) shows that such a lever rule does not
apply here. However, our  data are consistent with the more general
proposal  that the  bending rigidity behaves
smoothly both at $x_{\rm B} = 0$ and at $x_{\rm B} =1$.
Indeed, we conclude from the data in figure
\ref{usk}(b) that
\begin{equation}
1/\kappa \approx (1/\kappa_{\rm A}) ( 1 + \phi_{\rm AB} x_B )
\quad {\rm with} \quad \phi_{\rm AB} = 3.5  \pm 0.1
\end{equation} 
for small $x_{\rm B}$ and that
\begin{equation}
1/\kappa \approx (1/\kappa_{\rm B}) ( 1 + \phi_{\rm BA} x_A )
\quad {\rm with} \quad \phi_{\rm BA} = 1.0 \pm 0.2
\end{equation} 
for small $x_{\rm A}= 1 - x_{\rm B} $.

\begin{figure}[h]
\center
\psfrag{A}[br][br][1.]{$a_{in}\,  \pq{\rho^{2}}$}
\psfrag{xb}[ct][ct][1.]{$x_\tB$}
\includegraphics[height=5.5cm,width=8.cm]{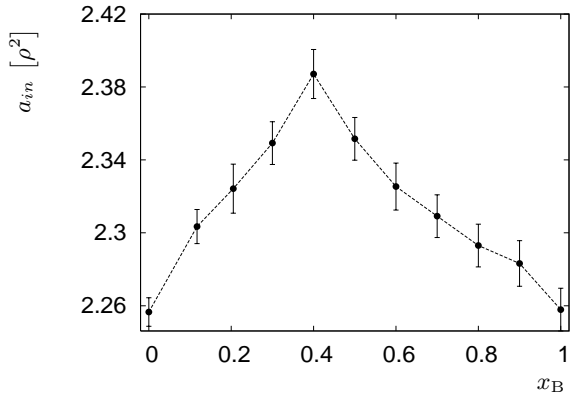}
\caption{Area per amphiphile $a_{in}=2A/N$ 
vs. the shorter amphiphile mole concentration $x_\tB$. The error bars represent standard errors obtained from the area fluctuations. The line 
is a guide to the eye. The units are given in brackets.}
\label{usa}
\end{figure}

We also measured the intrinsic bilayer area $A$, defined by
$A=\int_{A_p} \de x \de y \sqrt{1+(\nabla h)^2}$.
In Fig. \ref{usa}, the  intrinsic area per amphiphile, $a_{in}=2A/N$,
is plotted as a function of $x_\tB$ for tensionless bilayer states.
As shown in this figure, the intrinsic molecular area $a_{in}$ is again
found to be a nonmonotonic function of $x_\tB$:
for $x_\tB=0$, it has the value $a_{in}\simeq2.26\rho^2$,
but exhibits a maximum
at $x_\tB= 0.4$ with $a_{in}\simeq2.39\rho^2$.
The change in the  intrinsic area is obviously connected to the change
in the bending rigidity.
The decrease of the latter quantity corresponds to a more flexible bilayer
with larger shape fluctuations and a corresponding increase in the
intrinsic area.
Inspection of figure \ref{usa} shows that the pure A--bilayer has essentially the same
intrinsic molecular area as the pure B--bilayer, i.e., $a_{in}(x_\tB = 0)
\simeq a_{in}(x_\tB = 1) \simeq 2.26 \rho^2$. In contrast, the corresponding
{\em projected} molecular areas $a_{p}$ are found to be $a_{p}(x_\tB = 0)
= 2.10\rho^2$ and
$a_{p}(x_\tB = 1) = 1.96\rho^2$ which reflects the increased shape fluctuations for
$x_\tB = 1$.

\begin{figure}[h]
\center
\psfrag{sig}[br][br][1.]{$\gp\, \pq{\epsilon/\rho^{2}}$}
\psfrag{xb}[ct][ct][1.]{$x_\tB$}
\includegraphics[height=5.5cm,width=8.cm]{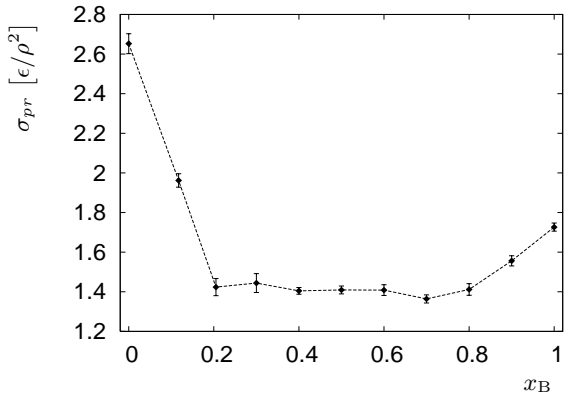}
\caption{Protrusion tension  $\gp$
vs. the shorter amphiphile mole concentration $x_\tB$. The error bars represent statistical errors obtained by the fit of the fluctuation spectrum data  to eq. (\ref{sq2}). The line is a guide to the eye. The units are given in brackets.}
\label{sigma_pr}
\end{figure}

Finally, the protrusion tension $\gp$ as introduced in (\ref{sq2})
also exhibits a nonmonotonic behavior as shown in Fig. \ref{sigma_pr}.
A decrease in the protrusion tension implies an increase
in the amplitude of
the local protrusions which become energetically more favorable.
As shown in Fig.  \ref{sigma_pr},  $\gp$ decreases with increasing $x_\tB$
up to
$x_\tB=0.2$, then stays essentially constant up to $x_\tB=0.8$, and finally
increases \
again up to $x_\tB =1$. This indicates
that  a small amount of shorter B--amphiphiles is
sufficient to roughen the bilayer surface, and further addition of these
amphiphiles has essentially no effect on the local protrusions until one
reaches another regime characterized by a bilayer of B--amphiphiles with
a few longer A--amphiphiles.
Comparing the results for bending rigidity,  intrinsic area and
protrusion tension, one concludes that the changes
in  the intrinsic area depend both on the
shape fluctuations and on the bilayer
roughness arising from the local mismatch between the two types of amphiphiles.

The bending rigidity of amphiphilic bilayers has been measured in
many experiments: while phospholipid membranes \cite{LB,EDS,Ev,BHB,MG} are
characterized by a bending rigidity of the order of tens of
$k_BT$, lamellar and fluid microemulsion phases composed of single chain
surfactants,
short chain alcohols, oil and water \cite{MML,SRS,SSR,LL} exhibit
a bending rigidity of the order of $k_BT$.
The bilayer studied in our Molecular Dynamics simulations are composed
of single--chain amphiphiles  and
have bending rigidities which are of the same order of magnitude as  those
studied in the second group of experiments.
In Refs. \cite{MML,SSR,SRS},
the response of lamellar systems to the insertion of shorter cosurfactants
was studied experimentally.
In these systems, the main surfactant was sodium dodecyl sulfate (SDS),
which is
a single chain surfactant, while the cosurfactants were alcohol
molecules: it was found that the bending rigidity decreases
as the fraction of shorter chain cosurfactant increases.
In Ref. \cite{SSR}, short surfactant (pentanol) molecules were added to
bilayers composed of two--chain amphiphiles (DMPC)  and a decrease of the
bending rigidity
was observed.
Two--component   bilayer membranes were also studied
theoretically by Szleifer et al. \cite{szkr} using a simple mean field
theory to treat
chain packing. In the framework of these theories, which are based on
smooth bending
configurations and, thus, neglect protrusions, the longer amphiphiles obtain
more
configurational freedom as the shorter amphiphiles are added which leads to an
increased flexibility of the bilayers. The behavior of the bending rigidity as
found in our simulations for small $x_\tB$ is consistent with such a mechanism.
However, we also find that both the bending rigidity and the protrusion tension
{\em decrease} as one adds long A--amphiphiles to a bilayer of short
B--amphiphiles for large values of $x_\tB$,
see Figs. \ref{usk} and \ref{sigma_pr}. In this latter
regime, the
addition of the A--amphiphiles induces more protrusions which act to
reduce the
bending rigidity \cite{LG}. 

In conclusion, we have presented the results of Molecular Dynamics simulations
for a  coarse grain model of  two-component bilayers.
We show that the functional dependence of the inverse bending rigidity
on the membrane composition does not follow a simple lever rule.
In contrast, we find that both the addition of short amphiphiles
to a bilayer of long ones and the addition of long amphiphiles to a bilayer
of short ones
leads to an increase in molecular protrusions and to a decrease in the
bending rigidity.
We also observe a nonmonotonic functional dependence of the intrinsic area
on the membrane composition which
is strongly correlated with the behavior of both local protrusions and
long--wavelength shape fluctuations.
The simulation approach described here can be  extended to
bilayer membranes with three components which have recently been shown to
lead to coexisting membrane domains or `rafts' and to more realistic models
of biological membranes which contain membrane proteins.\\

\end{document}